# Imaging of isotope diffusion using atomic-scale vibrational spectroscopy

*Ryosuke Senga[1], Yung-Chang Lin[1], Shigeyuki Morishita[2], Ryuichi Kato[1], Takatoshi Yamada[1], Masataka Hasegawa[1], and *Kazu Suenaga[3]

[1]*Nanomaterials Research Institute, National Institute of Advanced Industrial Science and Technology (AIST); 1-1-1 Higashi, Tsukuba, Ibaraki 305-8565 Japan*
[2]*JEOL Ltd.; 3-1-2 Musashino, Akishima, Tokyo 196-8558, Japan*
[3]*The Institute of Scientific and Industrial Research (ISIR), Osaka University; Mihogaoka 8-1, Ibaraki, Osaka 567-0047, Japan*

**Abstract**

The spatial resolutions of even the most sensitive isotope analysis techniques based on light or ion probes are limited to a few hundred nanometres. Although vibration spectroscopy using electron probes has achieved higher spatial resolution[1–3], the detection of isotopes at the atomic level[4] has been challenging so far. Here we show the unambiguous isotopic imaging of $^{12}$C carbon atoms embedded in $^{13}$C graphene and the monitoring of their self-diffusion via atomic-level vibrational spectroscopy. We first grow a domain of $^{12}$C carbon atoms in a pre-existing crack of $^{13}$C graphene, which is then annealed at 600 °C for several hours. Using scanning transmission electron microscopy–electron energy loss spectroscopy, we obtain an isotope map that confirms the segregation of $^{12}$C atoms that diffused rapidly. The map also indicates that the graphene layer becomes isotopically homogeneous over 100-nanometre regions after 2 hours. Our results demonstrate the high mobility of carbon atoms during growth and annealing via self-diffusion. This imaging technique can provide a fundamental methodology for nanoisotope engineering and monitoring, which will aid in the creation of isotope labels and tracing at the nanoscale.

**Main text**

Isotope analysis techniques such as mass spectrometry, nuclear magnetic resonance analysis, infrared spectroscopy, and Raman spectroscopy have been used in various applications, such as tracing of biological and chemical reactions[5–7], environmental surveys[8], and age estimation of minerals[9]. The spatial resolutions of these techniques are limited to a few hundred nanometres, even for methods that use sensitive probes such as Raman spectroscopy and secondary ion mass spectrometry. Thus, contemporary isotope analysis techniques are based mainly on macroscopic phenomena. Further microscopic applications such as atom-by-atom tracking of biochemical reactions with isotopic molecular labels, in-situ observations of material growth and diffusion processes using isotopic reaction gases, and non-destructive isotope analysis of micro-samples such as nanofossils, artefacts, and space minerals, are expected to lead to breakthroughs in a wide range of fields. To realise such nanoscale isotope engineering, it is essential to develop an isotope detection technique with sub-nanometre spatial resolution.

Scanning transmission electron microscopy (STEM) provides atomic-resolution images that reflect the electrostatic potential of atoms. However, STEM cannot distinguish isotopes, except for slight differences in the knock-on damage ratio[10]. Recently, the energy range of infrared spectroscopy has been realised using electron energy loss spectroscopy (EELS) with a monochromatised electron source, which enables access to the lattice vibrations reflecting atomic weights[1–3,11,12]. Hachtel et al. successfully distinguished isotope-labelled biomolecules by scanning samples in aloof geometry[4]. However, the EELS signal delocalisation deteriorates the spatial resolution, because it employs dipole scattering in the bright-field geometry[4]. On the contrary, phonon measurements of single silicon atoms in graphene[1] and phonon modulation measurements at the grain boundaries of nanoparticles[2] have been reported using high-spatial-resolution vibrational spectroscopy, based on impact scattering in the dark-field geometry in EELS mode.

In this study, we focused on isotope analysis in graphene by employing vibrational spectroscopy in STEM. In previous studies, graphene has been doped with different isotopes to clarify its growth mechanisms[6,7] and to fabricate phonon devices with high thermoelectric efficiency[13]. However, these experiments were conducted on the scale of a few hundred nanometres to a few micrometres, and there have been no reports on the spatial distributions of isotopes below the nanoscale. Here, we demonstrate in-situ isotopic nanodomain growth within single-layer graphene via transmission electron microscopy (TEM) and report for the first time the successful detection of carbon isotopes at sub-nanometre spatial resolution, by using dark-field vibrational spectroscopy with a monochromatised electron beam. The proposed method was also applied to track the isotopic carbon atoms in graphene to investigate the self-diffusion of carbon atoms under heating conditions. These results are a proof-of-principle demonstration of nanoscale isotope monitoring using electron microscopy.

To monitor the carbon isotope on the nanometre scale, $^{12}$C graphene should be distinguished from $^{13}$C graphene via the EEL spectra. Because graphene is infrared-inactive, no phonon signals can be detected in bright-field EEL spectra. Thus, we chose a dark-field condition, as shown in Figs. 1a, which yields relatively strong signals[12]. Moreover, this approach reduces the EELS signal delocalisation due to the impact scattering geometry. The detailed experimental conditions, including the electron beam geometry, are shown in Figure 1a-c and the Methods section. Figure 1d compares the reference spectra of a single layer graphene composed of $^{12}$C and $^{13}$C in the dark-field condition. Because a wide range of momentum transfer $q$ including a full third Brillouin zone was integrated in this condition, it was not straightforward to separate every vibrational mode from the spectrum. However, two prominent peaks could be identified clearly: the peak at higher energy (H-peak) corresponds to the optical vibration modes (LO+TO), and the broad peak at lower energy (L-peak) involving mainly the acoustic vibration modes (LA+TA+ZA). The peak positions extracted by fitting with the Voigt functions provide shifts of approximately 3 meV for the L-peak and 8 meV for the H-peak between $^{12}$C and $^{13}$C dominant graphene (Extended Data Table S1). Note that both spectra were obtained from reference samples of $^{13}$C and $^{12}$C graphene with 99% isotopic purity. The L-peak is difficult to interpret because it contains several vibrational modes with dynamic dispersion relations and appears on the tail of a quasi-elastic scattering peak. The H-peak is composed of only optical modes with relatively flat dispersion relations[12]. Therefore, $^{12}$C and $^{13}$C can be clearly distinguished by analysing the position of the H-peak. The peak shift of 8 meV obtained here is consistent with the shift of the phonon density of states (PDOS) between $^{12}$C and $^{13}$C graphene calculated by density function perturbation theory (DFPT) (Extended Data Fig. 1).

We prepared domains with different carbon isotopes by growing graphene at the cracks in TEM, based on the method developed by Liu *et al.*[14]. In this method, residual gas is used in the TEM chamber as a carbon source for the extensional growth of a graphene layer on the edges under STEM observation (Fig. 2a). Figure 2c shows a TEM image of a typical crack in single-layer $^{13}$C graphene before extensional growth, where a silicon nanoparticle remaining from the sample preparation process becomes located at the end of the crack. The silicon nanoparticle plays an essential role as a catalyst in the graphene growth process, and it is dominated by Si-C bonds according to the silicon L-edge energy-loss near-edge structure before the growth occurs (Fig. 2b). When the sample temperature was set to 650 °C and the electron beam was focused onto an area of 80–300 nm$^2$ in the TEM mode, the silicon nanoparticles started to move along the edge while changing the shape (Figs. 2d–h). Then, the crack was closed by embedding with a newly formed graphene domain. The new domain with a width of 1–2 nm and a length of 10 nm (approximately 400 carbon atoms) covering the original crack was formed within 3 min. The cumulative electron dose in the process was $1.3 \times 10^9$ e$^-$/nm$^2$. The growth rate of carbon atoms per electron beam density was $3.1 \times 10^{-7}$ atoms·nm$^2$/e$^-$, which is approximately equal to the value reported in Ref. [14]. TEM shows that these newly grown domains mainly comprise $^{12}$C, which is the main component of the residual carbohydrate gas[14].

The newly formed domain exhibited prolific non-hexagonal defects such as pentagon, heptagon, and octagon rings in the early growth stage (Fig. 2d). These defects are repaired as the growth proceeds. Interestingly, the generated hole depicted in Fig. 2g was not repaired, because there was no silicon catalyst, and it expanded during the process. In addition, the $L$-edge fine structure of Si nanoparticles after the growth is different from that at the beginning of the experiment (Fig. 2b). The lowest energy edge ($L_{2,3}$ edge) before the growth has a broad peak at approximately 105 eV, whereas the $L_{2,3}$ edge after the growth is formed by a sharp peak at 108 eV and a shoulder at the lower energy side. These features are consistent with the Si $L$-edge in SiC[15] and SiO$_2$[16], respectively. This can be attributed to the oxidation of Si nanoparticles by the residual gas in the TEM chamber during the growth process. In fact, because the oxidation process was observed in our experiments with successful growth, including the cases in Figs. 2 and 3, the electronic state of silicon can be regarded as an essential factor contributing to graphene growth.

After the growth, the sample temperature was lowered to 500 °C, and the TEM mode was switched to the STEM mode with the dark-field EELS condition to obtain vibrational spectra from the newly formed graphene domains. Figure 3a shows a TEM image after isotope domain growth of $^{12}$C carbon atoms at the crack of $^{13}$C graphene at the same position as shown in Fig. 2. Then, the EELS 2D image-spectra were obtained from the same region (Figs. 3b and c). Each spectrum was measured at an exposure time of 0.5 s with a pixel size of 0.35 nm × 0.35 nm, which is slightly larger than the probe size and covers approximately four carbon atoms, as shown in Extended Data Fig. 2. All obtained spectra were processed by the principle component analysis (PCA) algorithm using 50 components, which reduced the noise without losing features[17]. Then, the spectra were fitted with Voigt functions to identify the H-peak position in each pixel. Figure 3d presents plots of the peak positions using a line scan along the yellow line in Fig. 3a obtained using the same process. The variations of the peak positions can be classified into three categories: (1) a region mainly composed of $^{13}$C (165–170 meV), (2) a region mainly composed of $^{12}$C (175–180 meV), and (3) an intermediate region containing both $^{12}$C and $^{13}$C (170–175 meV). Figure 3e depicts the typical spectra for the three cases.

The colour coordination map of the H-peak position shows that the $^{12}$C signal is dominant in the newly formed graphene domain (Fig. 3c). This result suggests that the in-situ growth of graphene in TEM is not due to the decomposition and rearrangement of the existing carbon layer, but rather due to the residual gas in the TEM chamber as the carbon source. Indeed, the presence of residual gases such as hydrocarbons in the TEM chamber was confirmed in-situ by quadrupole mass spectroscopy equipped with TEM (Extended Data Fig. 4).

It seems that incorporation of $^{12}$C atoms can also occur without silicon atoms. Figure 3c reveals the existence of $^{12}$C around another hole (the white arrows). This hole was not repaired, but rather expanded during the growth as there were no silicon atoms near it. The $^{12}$C domain around the hole suggests that the incorporation of $^{12}$C atoms from the residual gas and their diffusion took place under electron beam irradiation at the edge of the hole. This possible self-diffusion of carbon atoms during the observation may explain why the $^{12}$C region is not fully localised to the original crack area.

Figure 4 shows the diffusion of $^{12}$C graphene grown along different cracks of $^{13}$C graphene. Before the growth, there was a crack more than 50 nm long and 10–20 nm wide with a silicon nanoparticle at its end (Fig. 4a). At this point, isotope mapping by STEM-EELS showed that the graphene around the crack was mostly composed of $^{13}$C (Fig. 4d). The $^{12}$C and $^{13}$C mixed regions found in a small part of the sample may have been present from the beginning of the CVD process. A $^{12}$C graphene domain was grown at the crack by using the same method shown in Fig. 2. After electron beam irradiation for 5 min at 650 °C, the crack depicted in Fig. 4a was fully filled by the newly grown graphene domain (Fig. 2b). The isotope mapping measured immediately after the growth shows that the $^{12}$C region is localised around the newly grown area (Fig. 4e). However, even at this point, the newly grown domain is not a homogeneous $^{12}$C domain but is comprised partly of $^{13}$C.

Then, the sample was heated at 600 °C for 2 h. By comparing the atomic structure of the newly grown domain before (insets in Fig. 4b) and after heating (insets in Fig. 4c), it was found that the number of defects in the same position was lower after heating because of structural relaxation and annealing. In addition, the slightly extended second and third layers that formed during the heating process can be seen in the lower part of Fig. 4c. Isotope mapping at this stage shows that the $^{12}$C region seen in Fig. 4e diffuses almost completely, and only small mixed regions of $^{12}$C and $^{13}$C remain (Fig. 4f).

The self-diffusion coefficient was simply estimated from the equation $D = x^2/4t$, where $x$ and $t$ are the mean distance of diffused atoms and diffusing time, respectively.[18] Although it is difficult to measure the self-diffusion coefficient accurately from the present results, it is expected to be at least larger than $3 \times 10^{-20}$ m$^2$/s, based on the fact that $^{12}$C completely diffused to the outside of the measurement area of 60 nm square after 2 h of heating. This rate is nearly two orders of magnitude faster than the previously reported[18] diffusion rate of impurity platinum atoms on graphene at the same temperature (D = $4 \times 10^{-22}$–$1 \times 10^{-21}$ m$^2$/s). Such faster self-diffusion is expected to proceed through the direct exchange of carbon atoms along the defect-free hexagonal rings[19], as well as the diffusion and repair of defects[20] created during the growth process.

To rule out the effect of atomic defects on the H-peak position, we performed a set of experiments to corroborate the vibrational spectra from highly defective areas in monoisotopic samples. Non-hexagonal carbon networks, such as a boundary of $^{13}$C graphene (Extended Data Fig. 9) and a $^{12}$C graphene domain, grew at the crack in the $^{12}$C graphene (Extended Data Fig. 10). In these experiments, the assigned H-peak position does not show any considerable change. With further improvements in energy and momentum resolutions, the effect of defects on the vibration mode could be quantitatively characterized in future studies, which is crucial for controlling the thermoelectric and optoelectric properties of materials.

The noise level for $^{13}$C/$^{12}$C graphene isotope maps has been statistically analysed. The variation of the measured vibrational peak positions for isotopically pure $^{12}$C/$^{13}$C graphene is ranging from 2.0 meV to 3.0 meV, which confirms the high signal-to-noise ratio of $^{12}$C atom detection from $^{13}$C graphene (Extended Data Fig. 8). To realise much higher confidence level for monatomic isotope detection, it is necessary to improve the signal-to-noise ratio of the spectra by reducing both the real-space and momentum-space integration effects probably with the combination of a coherent high-brightness electron source and a further stability-improved monochromator. Further improvements in the energy resolution are also highly desired for the application of this method to heavier elements such as Cl, P, and S, which are widely used as isotopic labels in biological specimens.

Theoretical aspects also should be taken into account. Recently, localised spectral changes have been reported from theoretical calculations using hexagonal boron nitride, such as molecules containing isotopic impurities[21]. Theoretical studies of such local isotope effects in periodic and non periodic structures should be further developed together with experiments.

In this study, we succeeded in performing isotope mapping of electron beam-induced growth of graphene at a subnanometre resolution by means of vibrational spectroscopy using dark-field STEM-EELS. This imaging technique can act as a fundamental method for nanoisotope engineering and monitoring, which will aid in the creation of isotope labels and tracing on the nanoscale. Further improvements in energy and spatial resolution and sensitivity may push isotope measurements to the monatomic scale and enable applications such as the tracking of single-isotope atomic labels in biological and chemical reactions.

**Acknowledgements**:

This work was supported by JST-PRESTO (JPMJPR2009), JST-CREST (JPMJCR20B1, JPMJCR1993), JSPS KAKENHI (16H06333, 21H05235), ER-C "MORE-TEM" and NEDO (JPNP16010) projects.

**Author contributions**

RS and KS designed the experiments. RS performed the in-situ graphene growth in TEM. RS and SM performed EEL spectroscopy. RS analysed the data. RS, YCL, and RK prepared the samples on the TEM grid. RK, TY, and MH prepared the samples by CVD. RS and KS co-wrote the paper. All authors commented on the manuscript.

**Competing interests**

The authors declare no competing financial interests.

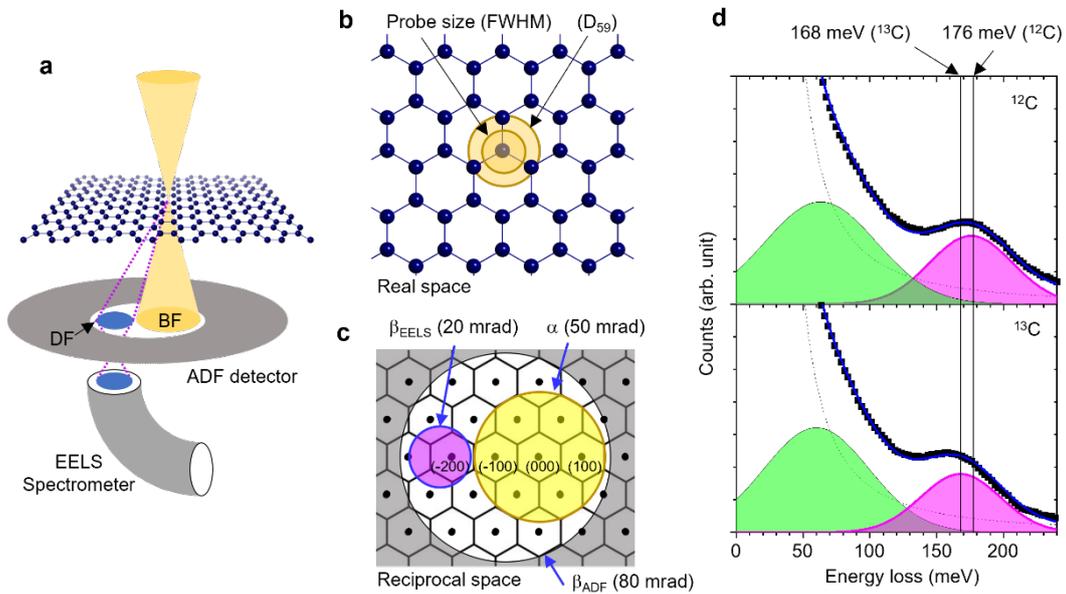

**Figure 1 Comparison of $^{12}$C and $^{13}$C graphene by dark-field EELS.**
**a**. Electron beam geometry of dark-field EELS. **b**, **c**. Probe size in the real space and its corresponding geometry at the reciprocal space. The convergence semi-angle α of the incident beam is 50 mrad, which corresponds to the bright field shown by the yellow circle in **c**. The EELS collection aperture was placed off-centre (purple ray path in Fig. 1a) and covered an area of β = 20 mrad in the diffraction plane, as shown in **c**. Both the EELS aperture and bright field were placed within the inner angle of the annular dark-field (ADF) detector. **d**, Vibrational spectra of $^{12}$C and $^{13}$C graphene acquired by dark-field EELS, where the spectra obtained by scanning a 20 nm × 20 nm region with few defects were integrated. This EELS condition provides two components in the vibrational excitation. The lower energy component at approximately 60 meV consists of the TA, ZA, LA, and ZO modes (L-peak). The higher energy component at approximately 170 meV consists of the TO and LO modes (H-peak). All peaks, including the elastic scattering peaks indicated by the broken lines, were fitted by Voigt functions. The raw data and fitting curve including all components are shown as black squares and a solid blue line, respectively. The frequencies of the optical vibrational mode for $^{12}$C and $^{13}$C graphene extracted by the fitting are 176 meV and 168 meV, respectively. Extended Data Table S1 lists the fitting parameters, and Extended Data Fig. 6 shows the spectra for two layers. In the case of a homogeneous sample, the signal is almost identical regardless of the position of the EELS collection aperture in the dark field, that is, the in-plane orientation of the graphene (Extended Data Fig. 3)

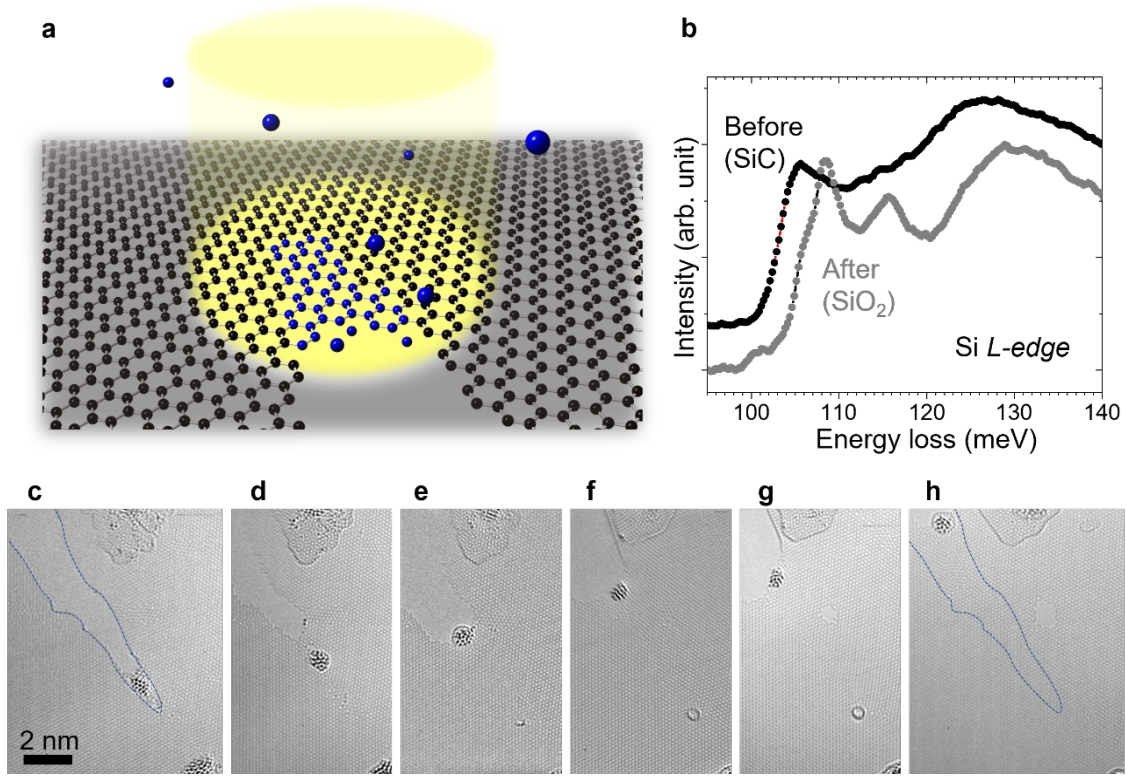

**Figure 2 In-situ growth of isotope nanodomains.**

**a**. Schematic of graphene growth in TEM. **b**. *L*-edge fine structures of a silicon nanoparticle before and after the growth process. Note that these spectra were obtained from a different experiment but through the same process. **c**–**h**. TEM images of the actual growth process. The blue dashed line indicates the crack position before the growth process. The non-hexagonal defects on the bottom of the silicon particle in **d** are almost repaired in **e**. In **g**, a hole is created, and it is not repaired as there is no silicon catalyst around the hole.

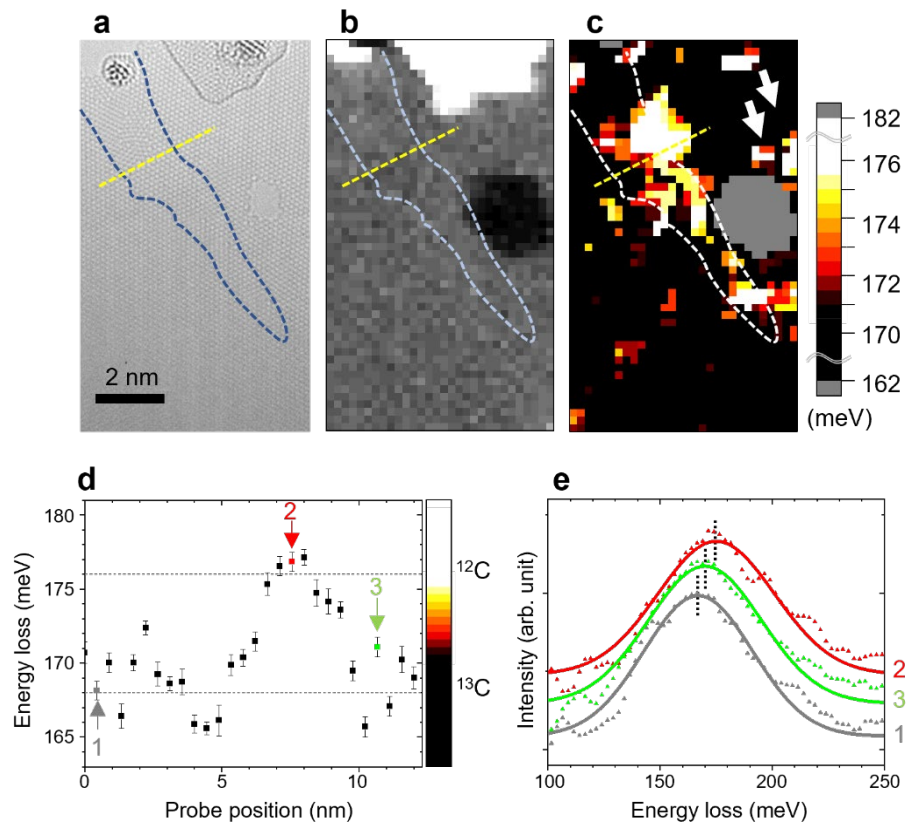

**Figure 3 Isotopic mapping of single-layer graphene.**
**a**. TEM image after isotope nanodomain growth (same as Fig. 2h). **b**. ADF image obtained by performing a STEM-EELS 2D scan at the same position as in **a**. **c**. Colour map showing the position of the high-energy peak. The original 2D spectrum image for this colour map was acquired from 51 × 58 pixels, where the exposure time of a single pixel (0.35 nm square) was 0.5 s. The peak positions of the optical vibration modes were obtained by fitting the PCA-processed spectrum with a Voigt function. The colour map was noise-processed using a 3 × 3 median filter. Extended Data Fig. 5 provides the unprocessed image. **d**. Energy distribution of the high-energy peak obtained by a line scan along the yellow line in **a**, where the spectra were collected from 0.29 nm × 28 pixels with an exposure time of 1.0 s for a single pixel. **c** and **d** are from separate experiments acquired in succession. **e.** Three typical spectra picked up from the plot of **d**. The numbers 1, 2, and 3 indicate $^{13}C$, $^{12}C$, and the intermediate region, respectively. The triangular dots present the PCA processed data, and the solid lines show the fitting curves with Voigt functions.

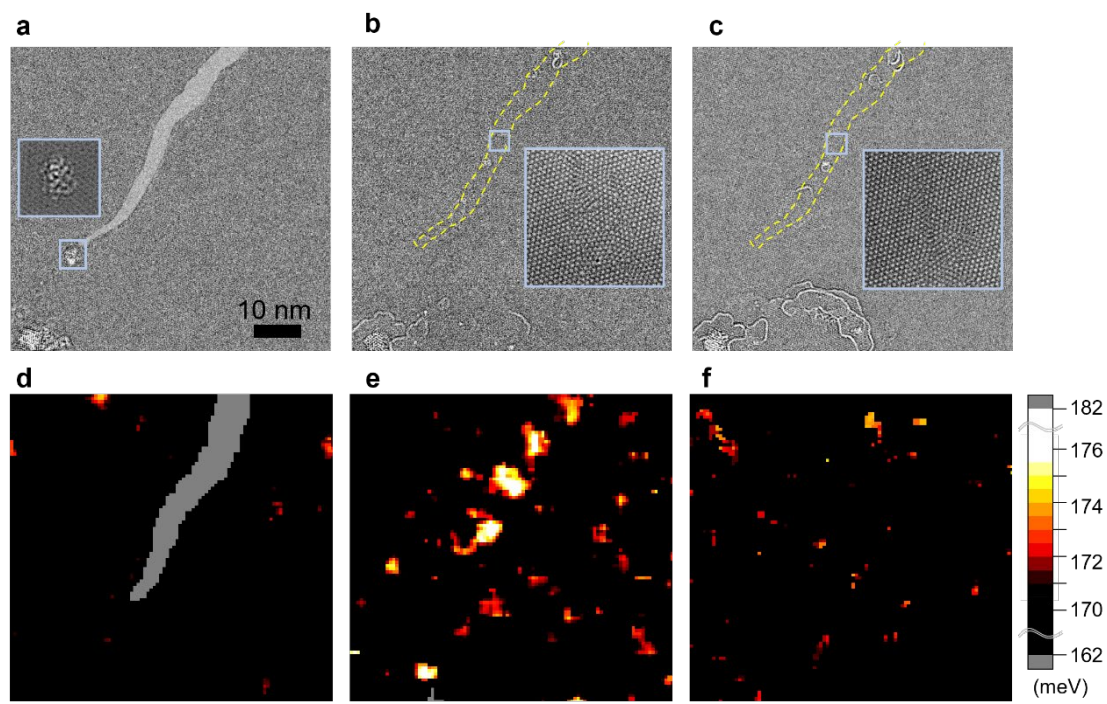

**Figure 4 Self-diffusion of carbon atoms in graphene.**
**a**, TEM image of $^{13}$C graphene before isotope nanodomain growth, showing a silicon nanoparticle at the end of the crack. **b**, TEM images of $^{13}$C graphene immediately after isotope nanodomain growth and **c**, the same location after heating at 600 °C for 2 h. Immediately after the growth process, non-hexagonal structures are randomly formed (inset in **b**). After heating, however, such defects are reduced through the structural relaxation and diffusion of the defects (inset in c). **d–f**, Colour maps of the high-energy peak positions corresponding to **a–c**. The 2D spectrum images were acquired in a 130 × 130 pixel area by exposing a single pixel (0.45 nm × 0.45 nm) for 0.5 s. All processes were similar to those in Fig. 3**c**. Note that some carbonaceous materials were present on the surface of the graphene, as shown at the bottom left of the images, which also changed shape during the experiment. The peak shift was small in this region. Therefore, the isotope ratio in the carbonaceous materials is not discussed here. The $^{12}$C domains in **e** that appear to cluster is mainly due to the effect of noise filter and does not imply the spontaneous 12C segregation during the growth.

**Methods**

Firstly, $^{13}$C graphene was synthesised by chemical vapour deposition (CVD) using $^{13}$C methane gas as the carbon source, and $^{12}$C graphene was synthesised as a control by using normal methane gas. The purity of the $^{13}$C methane gas was 99%. The percentage of $^{12}$C in the normal methane gas was approximately 99%, based on the natural isotope ratios. Graphene was transferred to TEM grids and heated at 500 °C in the TEM prior to observation.

All experiments were performed using a TEM (JEOL Triple C#2) equipped with STEM and TEM DELTA-type Cs correctors with a monochromator. The growth of $^{12}$C graphene nanodomains at a crack site of $^{13}$C graphene was monitored in TEM mode at a temperature of 650 °C. In-situ TEM observations were conducted with a Gatan One View camera. For similar in-situ growth, the growth rate has been determined by the vacuum level in the column in TEM, i.e., the amount of residual gas[14]. Then, the amount of $^{13}$C in the newly grown graphene domains will be negligible because the growth rate is almost three orders of magnitude higher than that in the self-healing process[22].

Nanoscale vibration spectroscopy was performed in the STEM mode, in which the convergence semi-angle of the incident beam and the EELS collection semi-angle at the dark field were 50 mrad and 20 mrad, respectively. The electron-beam geometry in the dark-field EELS is shown in Fig. 1. The probe current was 34 pA. The probe size was estimated to be 2 Å according to the full width at half-maximum and 2.7 Å at D$_{59}$, indicating a diameter of 59% of the electron current within a probe. Note that the probe size can be further reduced to approximately 1 Å with the aforenoted beam convergence condition. However, we used a rather enlarged electron probe to enhance the current. All electron energy loss (EEL) spectra were acquired using the Gatan GIF spectrometer designed for low-voltage TEM. The full-width at half-maximum of the energy resolution of the electron probe in the bright field was 18 meV. The spectra, including full zero-loss peaks at both the bright and dark fields, are shown in Extended Data Fig. 7.

Phonon dispersion and phonon density of state of $^{12}$C and $^{13}$C graphene were calculated within DFPT in Quantum Espresso[23]. The pseudopotential was the ultrasoft type and used local-density approximation with Perdew-Zunger parameters. A 32 × 32 × 1 mesh was utilized for the k points. The cut-off energy for the expansion of the wavefunction in plane waves and charge density were set to 50 Ry and 200 Ry, respectively. The phonon dispersion was calculated using a dynamical matrix on a 12 × 12 × 1 mesh. The masses of $^{12}$C and $^{13}$C were set to 12 and 13, respectively.

**Method References**

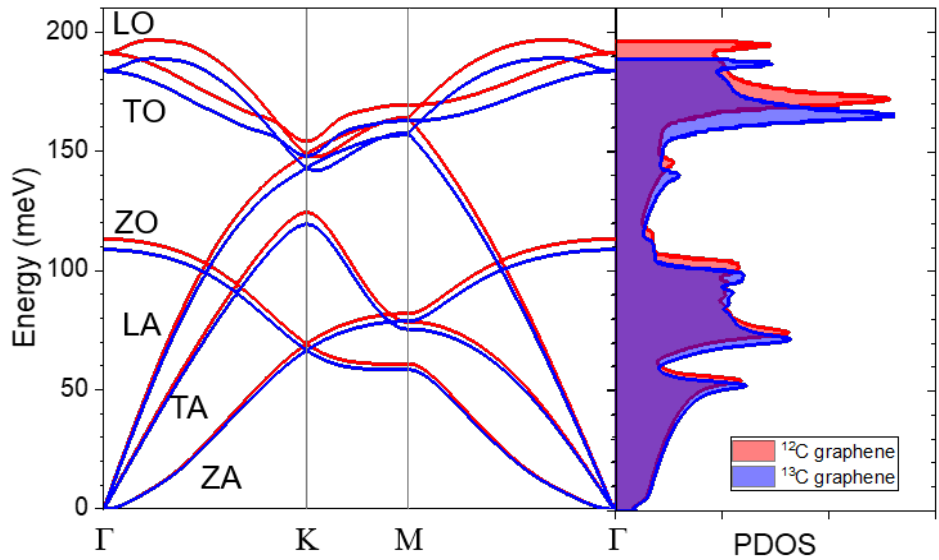

**Extended Data Fig. 1 Simulated phonon dispersion and phonon density of state (PDOS).**
The phonon dispersion (left) and PDOS (right) of $^{12}$C and $^{13}$C graphene obtained by DFPT calculations in Quantum Espresso. The energy difference between $^{12}$C and $^{13}$C graphene is 7.7 meV at the highest energy peak in PDOS corresponding to the LO mode and 6.7 meV at the second highest energy peak which is mainly contributed from the LO/TO mode. The interatomic force constants calculated from DFPT is, for instance, 52.4 eV/Å$^2$ for the in-plane direction at Γ. This calculation was performed to estimate the energy shift of the PDOS of $^{12}$C and $^{13}$C with the simple model. To reproduce the experimental spectra, full calculation considering the charge modulation in the higher-order Brillouin Zone is necessary, although such calculation is beyond the scope of this study and is not addressed here.

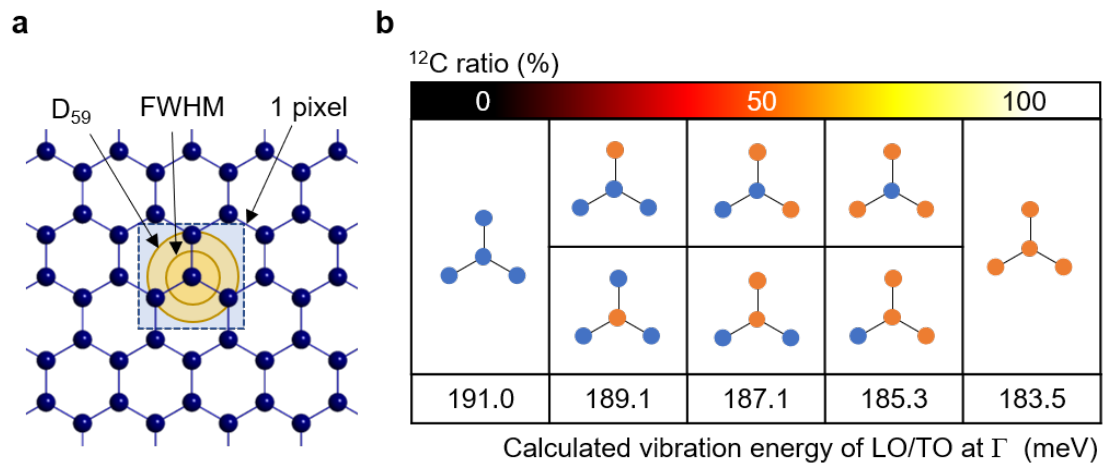

**Extended Data Fig. 2 Probe size and possible isotope configurations at the probed region.**

**a**, Comparison of probe and pixel size with graphene lattice. When the electron beam is fixed on an atom with our probe condition in which the probe size is increased to gain the current, the resulting spectrum contains the signal of the nearest three atoms in addition to that of the atom at the centre of the probe. Thus, a single spectrum roughly consists of the average signal of four atoms. The momentum space was also integrated up to 3.5 Å$^{-1}$. Therefore, the spatial resolution was influenced by the integration effect in both real and momentum space. Based on a study by Hage et al. using silicon single atoms on graphene, when the probe size is sufficiently small, a localised signal at the single atom level can be obtained in the dark-field EELS condition[1]. **B**, Isotope combinations of the four atoms. The colour distribution at the top in **b** corresponds to that used in Figs. 3 and 4. The corresponding vibrational energies of the LO/TO mode at Γ are calculated by DFPT and shown on the bottom line in **b**. When all four atoms are $^{12}$C or $^{13}$C, the energy difference of the H-peak between them is the largest and detectable with more than 90% of confidence level. If any one or two of the four atoms belong to different isotopes, the peak positions are between them. In this case, there are six possible configurations, as shown in **b**.

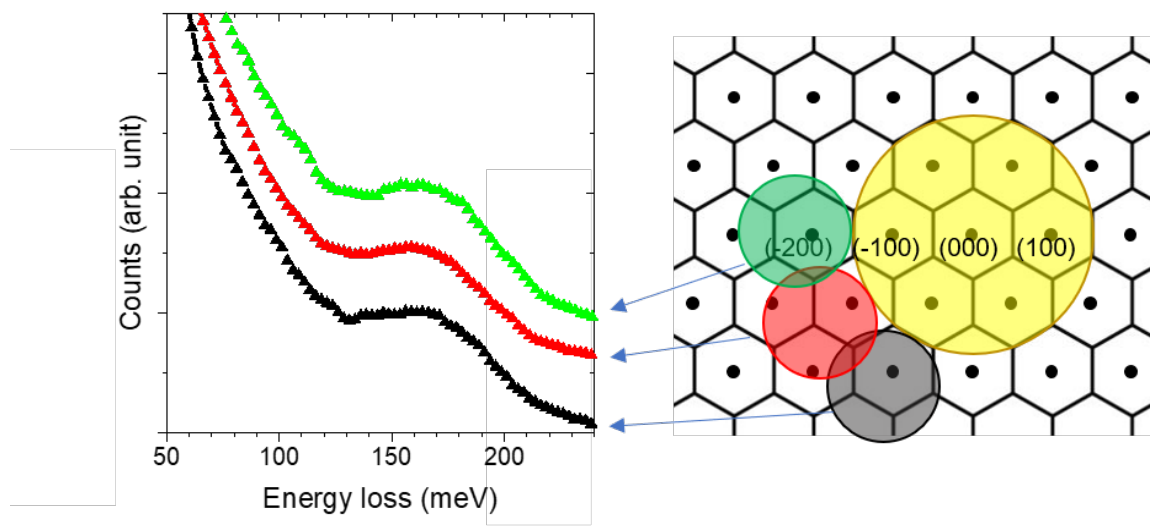

**Extended Data Fig. 3 Orientation dependence of vibrational spectra obtained by dark-field EELS.**

The signal is almost identical regardless of the position of the EELS collection aperture (green, red, and black); therefore, the in-plane orientation of graphene hardly affects the fitting parameters. The lattice defects in graphene also do not affect these parameters, except for the acoustic vibration mode in the lower peak, which is considered negligible.

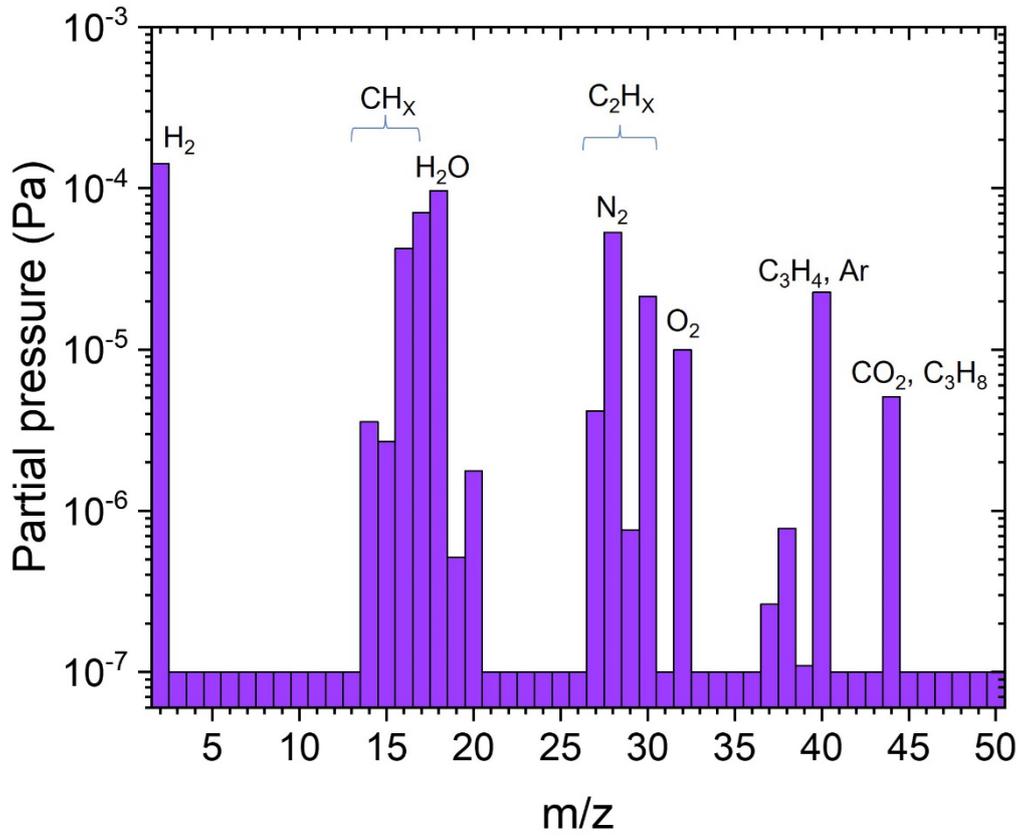

**Extended Data Fig. 4 Composition of residual gas in the TEM chamber measured by quadrupole mass spectroscopy.**
The residual gases include hydrocarbons such as $CH_x$ and $C_2H_x$.

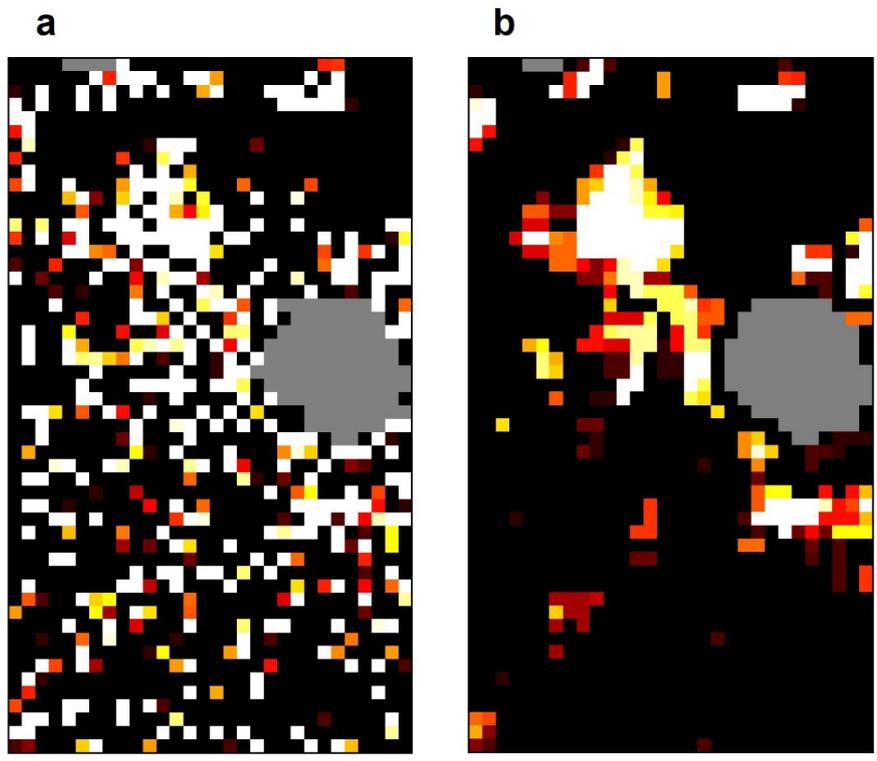

**Extended Data Fig. 5 Isotope colour maps before and after filtering.**
**a**,**b**, the isotope colour map in Fig. 3 before and after median filtering, respectively.

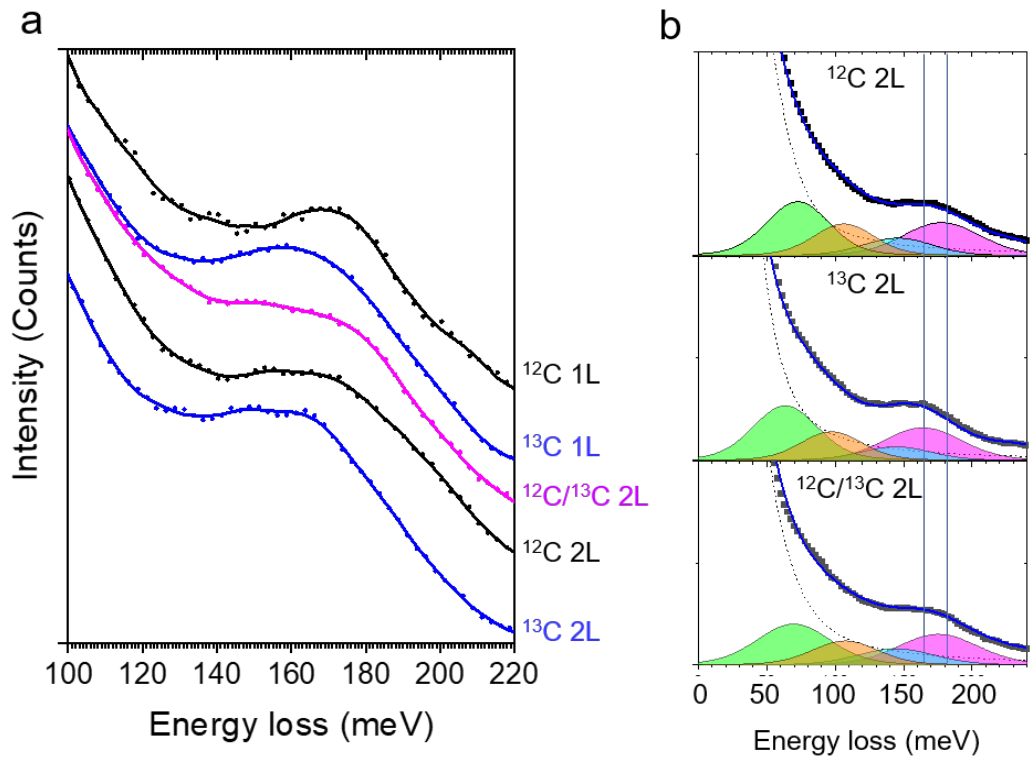

**Extended Data Fig. 6 EEL vibrational spectra for 1 layer (1L) and 2 layers (2L) graphene made of $^{12}$C, $^{13}$C and both stacked 2L ($^{12}$C/$^{13}$C).**
**a**, Comparison of the optical modes (H-peak) in the vibrational spectra of all the samples including 1L graphene shown in Fig. 1. The energy shift of approximately 7-8 meV between isotopes was found in 2L as in the case of 1L. The H-peak in 2L consists of two components due to possible contributions from the TO mode due to LO-TO splitting or the high-energy component of the LA mode. **b**, Line shape analysis of the vibration spectra obtained from 2L samples, where a rough component analysis by Voigt function is possible as in the case of 1L shown in Fig. 1. In this case, four components were used for the fittings: the L-peak, which is mainly the contribution of acoustic phonons, the two H-peaks mentioned above, and the peak that appears between them, which is considered to be the contribution of out-of-plane ZO mode.

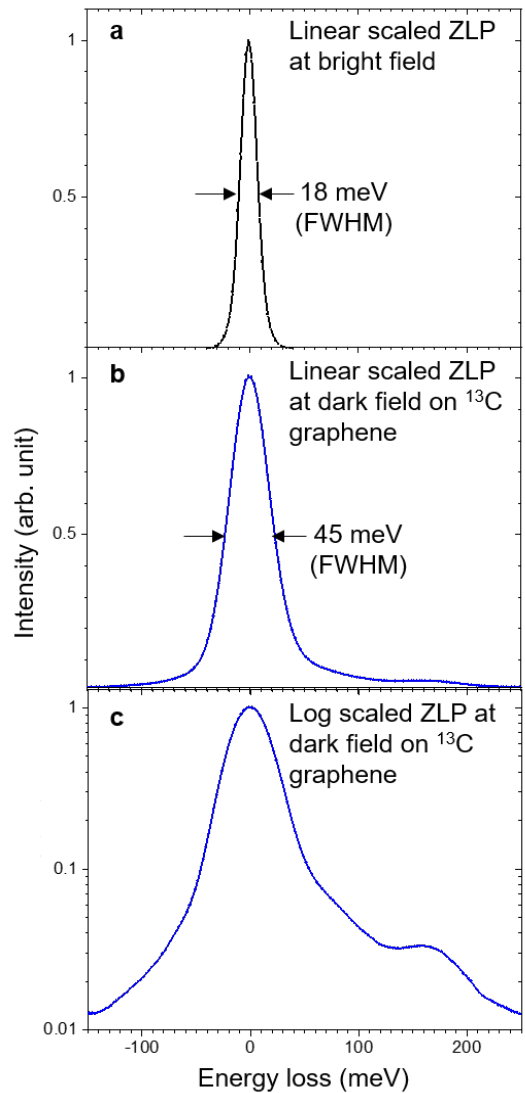

**Extended Data Fig. 7 EEL spectra including zero-loss peaks at bright-field and dark-field.**

a. EEL spectrum at bright field without samples. b, c, Linear- and log-scaled EEL spectra at dark field on $^{13}C$ graphene. The natural width of the zero-loss peak in the ideal on-axis condition has been shown. As an instrumental function, the zero-loss half-width is 18 meV, whereas it is 45 meV in the off-axis condition (with sample). The zero-loss peak in the off-axis is an ambiguous concept and cannot be detected without a sample.

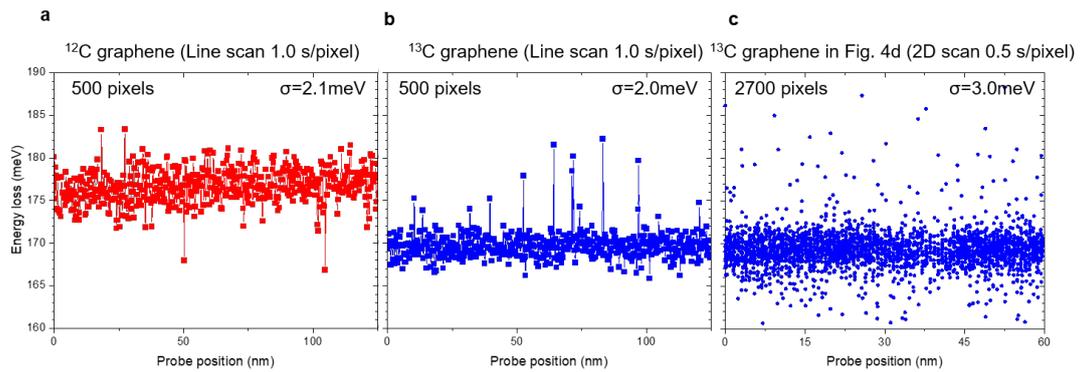

**Extended Data Fig. 8 Noise level quantifications of the measurements.**
**a**, **b**, The variation of the measured vibrational peak positions for line scans with an exposure time of 1 sec/pixel on $^{12}$C and $^{13}$C graphene, respectively. The standard deviations σ for 500 data points are 2.1 meV for $^{12}$C and 2.0 meV for $^{13}$C graphene, which is three-times smaller than the energy shift S between $^{12}$C and $^{13}$C graphene ~8 meV. Because the standard deviations of the measured points are based on fitting and contain errors, then the confidence intervals cannot be directly measured around those data points. This statical analysis, however, simply proves the high detection level of $^{12}$C atoms in case of 4 atoms and will give a standard for future experiments. Note that the purities of these samples are both over 99%, and thus, the peaking data points may be attributed to 1% of the isotopes in the samples. **c**, The variation of the measured vibrational peak positions on $^{13}$C graphene (corresponding to Fig. 4a) shows a σ of 3.0 meV at an exposure time of 0.5 s. Note that the data points at the crack are excluded.

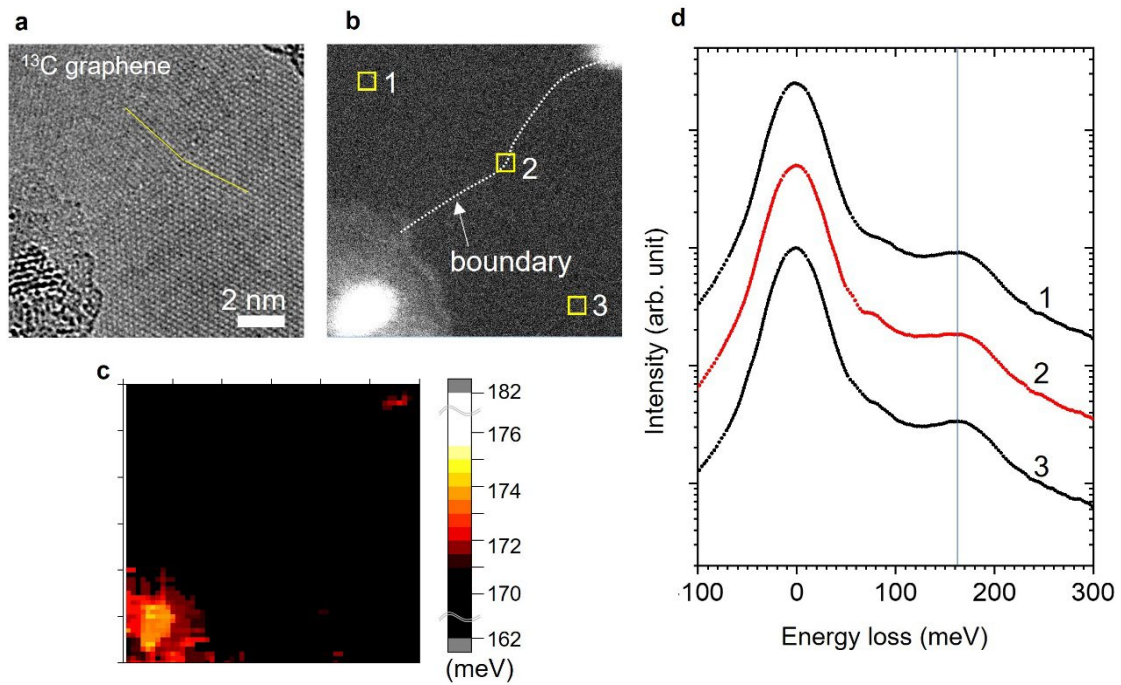

**Extended Data Fig. 9 2D vibrational spectroscopy across a grain boundary.**
**a**, TEM image of $^{13}$C graphene involving a grain boundary. The grain boundary extends from the top right to the bottom left of the image. The crystal orientation is rotated by approximately 16° across the grain boundary comprised of 5-7 membered rings, as indicated by the yellow lines. **b,** ADF image obtained by performing a STEM-EELS 2D scan at the same position as in **a**. **c**, Colour map of the high-energy peak positions corresponding to **b**. **d,** The EEL spectra taken from positions 1-3 in **b** are shown. The H-peaks in all three spectra are almost identical.

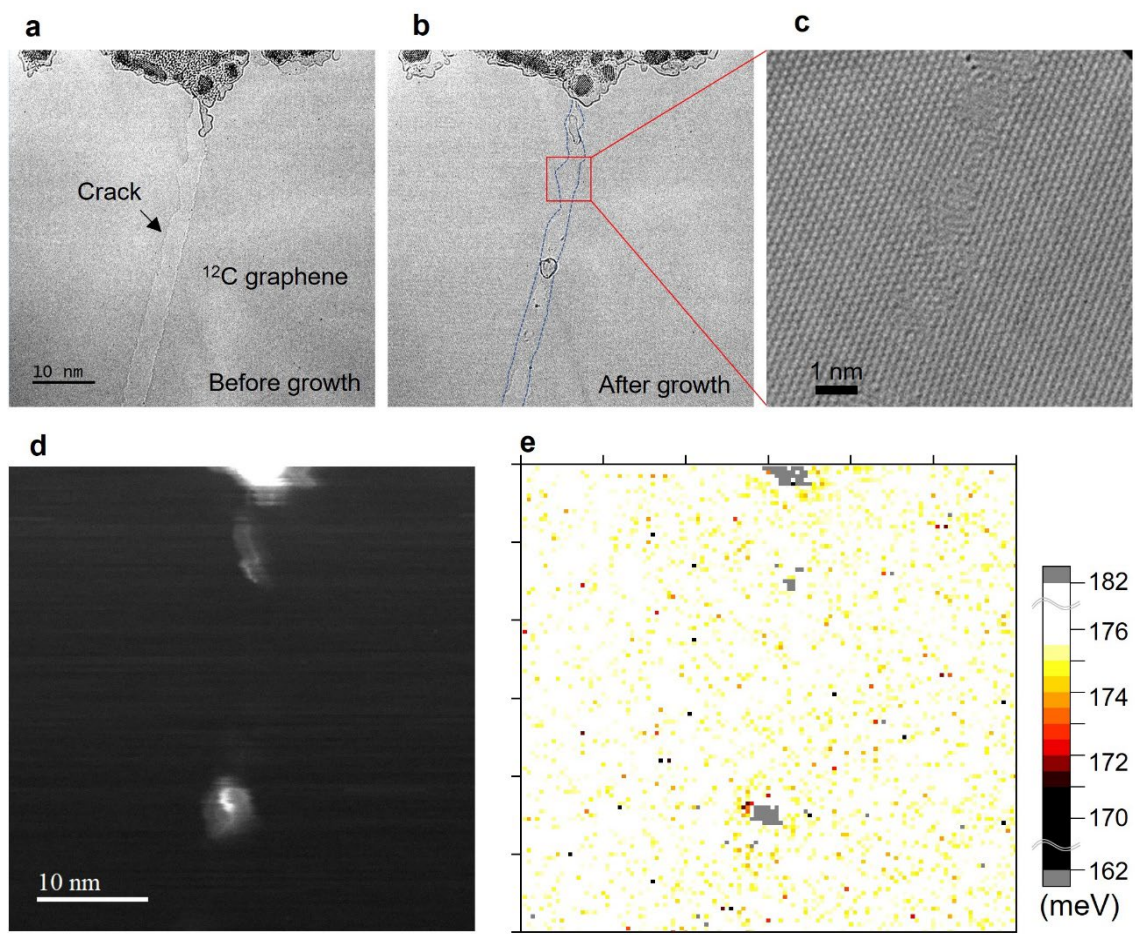

**Extended Data Fig. 10 H-peak position mapping on a crack of $^{12}$C graphene embedded by in-situ graphene growth in TEM.**

**a**, **b**, TEM image of initial $^{12}$C graphene including a crack and the same position after the nanodomain growth, respectively. The newly grown domain contains prolific defects involving 5-7 membered rings, as shown in **c**. **d**, ADF image obtained by performing a STEM-EELS 2D scan at the same position as in **b**. **e**, Colour map showing the position of the high-energy peak without noise filtering. The peak positions are almost uniform over the whole area, including the newly grown region.

**Extended Data Table S1** Fitted results of the vibrational spectra obtained from $^{12}$C and $^{13}$C graphene

| Peak position (meV) | Low-energy peak | High-energy peak |
| --- | --- | --- |
| $^{12}$C graphene | 63.1 (7.5) | 175.7 (4.9) |
| $^{13}$C graphene | 60.4 (4.5) | 167.5 (3.2) |